\documentclass[preprnt]{emulateapj}

\usepackage{epsfig}
\usepackage{portland}

\slugcomment{Submitted to the Astronomical Journal}
\shortauthors{Werchan \& Zaritsky}
\shorttitle{Structural Parameters of LMC Clusters}

\begin{document}

\title{The Star Clusters of the Large Magellanic Cloud: Structural Parameters}

\author{Felicia Werchan \& Dennis Zaritsky}

\affil{Steward Observatory, University of Arizona, 933
   North Cherry Avenue, Tucson, AZ 85721, USA}

\email{fwerchan@email.arizona.edu,\hfil \break dzaritsky@as.arizona.edu}

\begin{abstract}
We present and analyze the radial luminosity profiles of a sample of 1066 stellar clusters in the Large Magellanic Cloud. By design, this study closely follows the compilation by Hill \& Zaritsky of the structural parameters of stellar clusters in the Small Magellanic Cloud. 
Both King and Elson-Fall-Freeman (EFF) model profiles are fit to V-band surface brightness profiles measured from the Magellanic Cloud Photometric Survey images.
We tabulate the concentration, central surface brightness, tidal radii, 90\% enclosed luminosity radii ($r_{90}$), and local background luminosity density. Over two thirds of the clusters in the sample are adequately fit by one or both of these models. One notable and systematic exception, as in the SMC, are those clusters that lack a central brightness concentration, the ``ring" clusters.  While the bulk properties of the clusters are similar between the LMC and SMC populations, we find that the LMC lacks clusters that are as large, either in terms of core radii or $r_{90}$, as the largest in the SMC, perhaps a signature of larger tidal stresses in the LMC. 
\end{abstract}


\section{Introduction}

Gravitationally bound stellar clusters are the only known class of systems that, to the limit of our precision, generally consist of stars of the same age \citep[for some notable exceptions see][]{villanova, piotto, mackey08}.  As such
they play a central role in the development of our understanding of stellar evolution \citep[cf.][]{sandage}
and of the dynamical evolution of stellar systems \citep[cf.][]{king}. However, there is one significant shortcoming of the sample of Milky Way clusters as a testbed of models --- they are almost exclusively  old. While arguments continue regarding the exact spread in the ages of Milky Way globular clusters, the consensus is that they are mostly older than 10 Gyr \citep{salaris}. Fortunately, the stellar clusters in other local galaxies do not suffer the same cluster senility. 

For rich and young clusters in the Local Group, there exist age estimates based on the analysis of stellar color-magnitude diagrams \citep{p99, mackey04, glatt10}. However, the sample of clusters with such measurements and corresponding size, mass and other structural measurements is modest in
comparison to the total number of  nearby clusters. The Magellanic Clouds alone contain several thousand clusters. There are two complementary approaches to the study of these clusters. One can construct a high-quality, high-resolution sample of observations, usually using space-based data, of a limited number of clusters \citep{mackey03a, mackey03b, glatt}. These data produce the highest fidelity measurements of the structural
parameters and ages. Alternatively, one can measure these parameters more crudely, usually using ground-based data, but for a significantly larger sample of clusters. Because of our involvement in the Magellanic Clouds Photometric Survey \citep{zht, smc, lmc}, we have adopted the second approach. We have previously presented catalogs of stellar clusters in the Small Magellanic Cloud that quantified the structure \citep{hz} and age \citep{rz} for $\sim$ 200 clusters. Here we present the analogous catalog of structural properties for 1066 clusters in the Large Magellanic Cloud and compare those to the structural properties of the cluster populations of the SMC. We adopt an LMC distance of 50 kpc when converting to physical units. We briefly review the data and model fitting procedure in \S2 and 3, and discuss clusters properties in \S4.

\section{Data and Cluster Catalog}

The original $U$, $B$, $V$, and $I$ band images, from which the photometric stellar catalog was constructed, 
come from drift scans obtained with the Great Circle Camera \citep{zsb} mounted on the Swope 1m telescope at the Las Campanas Observatory in Chile for the Magellanic Clouds Photometric Survey \citep{zht}. In summary, these images have a pixel scale of 0.7 arcsec pixel$^{-1}$, exposure times ranging from about 4 to 5 minutes, and typical seeing of 1.5 arcsec. We obtained these data primarily to recover the star formation history of the Magellanic Clouds and that work is described by \cite{hz1,hz2}. To detect the candidate clusters presented here, we use the LMC photometric catalog \citep{lmc} and search for overdensities in that stellar distribution, as we did when studying the clusters of the Small Magellanic Cloud \citep{hz} and a portion of the LMC \citep{zht}. To search for clusters, we first bin the stellar density, for stars with $V < 20.5$, in units of 
10 arcsec pixels. Then, we produce a smoothed version of the stellar density on the sky using a median filter ten pixels across. By subtracting the latter from the former, localized concentrations are highlighted. We use SExtractor \citep{bertin} to identify objects, requiring 5 contiguous pixels that are above a 7$\sigma$ local threshold.

\begin{figure*}[]
\begin{center}
\plotone{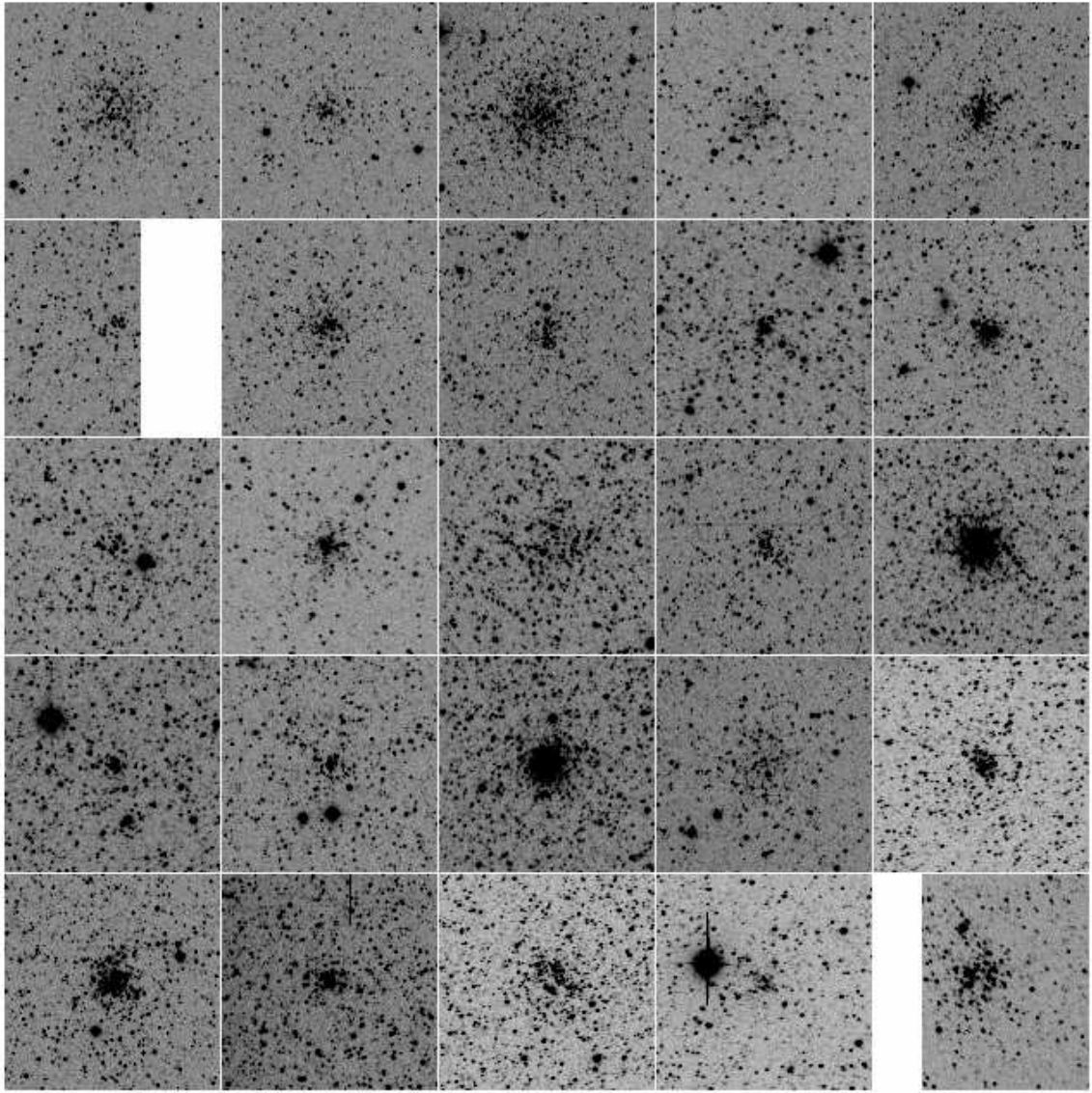}
\caption{$V$-band image mosaic of the first 25 LMC clusters in Table \ref{tab:Names and Positions}. Each image is the central subsection of the image used to fit the profiles shown in Figure \ref{fig:Sample Profiles} and spans 175 arcsec. Mosaics presenting all of the clusters in the sample
are available on-line.}
\label{fig:Sample Images}
\end{center}
\end{figure*}

We compile a list of 1225 candidate clusters. This list is certainly neither complete, nor devoid of false detections. Our aim is to err on the side of caution, so we visually examine all candidates and exclude those that are either visually non-distinct or strongly contaminated by nearby  sources (bright stars or nebulosity). Even in cases where these rejected candidates represent real clusters we would be unable to obtain reliable structural fits. We retain 1066 clusters and their coordinates and cross-identifications to the \cite{bica} catalog are presented in Table \ref{tab:Names and Positions}, while their images are presented in Figure \ref{fig:Sample Images}. The \cite{bica} catalog is an update of their earlier work in which they complement their collection of data from the literature with visual examination of photographic survey plates. As such, it is an extensive summary of 
the available data, but somewhat heterogeneous despite best efforts. That catalog includes all manner of extended objects, whereas we focus on stellar clusters and associations. Major and minor axis lengths are 
included, but profiles are not fit.
The catalog cross-identification is done matching to the nearest
source within 10 arcsec. Out of our total of 1066 clusters, we find matching designations in the literature for 682 (64\%).

\begin{deluxetable*}{cccc}
\tablecaption{Cluster Designation and Position}
\tablewidth{0pt}

\tablehead{
\colhead{Cluster} &
\colhead{$\alpha$} &
\colhead{$\delta$} &
\colhead{Alternate Designations}
}

\startdata
1	& $04^{h}\hspace{1 mm}32^{m}\hspace{1 mm}38.64^{s}$      & $-72^{\circ}\hspace{1 mm} 20^{\prime}\hspace{1 mm} 24.57^{\prime \prime}$ & SL4,LW4,KMHK7 \\
2	& $04^{h}\hspace{1 mm} 41^{m}\hspace{1 mm} 34.16^{s}$	& $-72^{\circ}\hspace{1 mm} 23^{\prime}\hspace{1 mm} 26.61^{\prime \prime}$ & LW27,KMHK39 \\
3	& $04^{h}\hspace{1 mm} 46^{m}\hspace{1 mm} 47.36^{s}$	& $-72^{\circ}\hspace{1 mm} 23^{\prime}\hspace{1 mm} 38.31^{\prime \prime}$ & SL37,LW61,KMHK98 \\
4	& $06^{h}\hspace{1 mm} 01^{m}\hspace{1 mm} 52.94^{s}$	& $-72^{\circ}\hspace{1 mm} 21^{\prime}\hspace{1 mm} 16.72^{\prime \prime}$ & SL826,LW363,KMHK1606 \\
5	& $04^{h}\hspace{1 mm} 45^{m}\hspace{1 mm} 54.31^{s}$	& $-72^{\circ}\hspace{1 mm} 21^{\prime}\hspace{1 mm} 07.30^{\prime \prime}$ & LW56e,KMHK83e \\
6	& $04^{h}\hspace{1 mm} 59^{m}\hspace{1 mm} 56.99^{s}$	& $-72^{\circ}\hspace{1 mm} 27^{\prime}\hspace{1 mm} 10.67^{\prime \prime}$ & \nodata \\
7	& $04^{h}\hspace{1 mm} 58^{m}\hspace{1 mm} 53.43^{s}$	& $-72^{\circ}\hspace{1 mm} 22^{\prime}\hspace{1 mm} 38.82^{\prime \prime}$ & SL157,LW109,KMHK398 \\
8	& $04^{h}\hspace{1 mm} 49^{m}\hspace{1 mm} 40.65^{s}$	& $-72^{\circ}\hspace{1 mm} 14^{\prime}\hspace{1 mm} 49.70^{\prime \prime}$ & LW69,KMHK137 \\
9	& $05^{h}\hspace{1 mm} 34^{m}\hspace{1 mm} 57.89^{s}$	& $-72^{\circ}\hspace{1 mm} 22^{\prime}\hspace{1 mm} 55.62^{\prime \prime}$ & LW247,KMHK1121 \\
10	& $05^{h}\hspace{1 mm} 09^{m}\hspace{1 mm} 00.94^{s}$	& $-72^{\circ}\hspace{1 mm} 21^{\prime}\hspace{1 mm} 41.80^{\prime \prime}$ & SL272,KMHK601 \\

\enddata
\tablecomments{The complete version of this Table is presented in the electronic edition of the Journal. The printed edition contains only a sample.}
\label{tab:Names and Positions}
\end{deluxetable*}

\begin{figure*}[htbp]
\begin{center}
\plotone{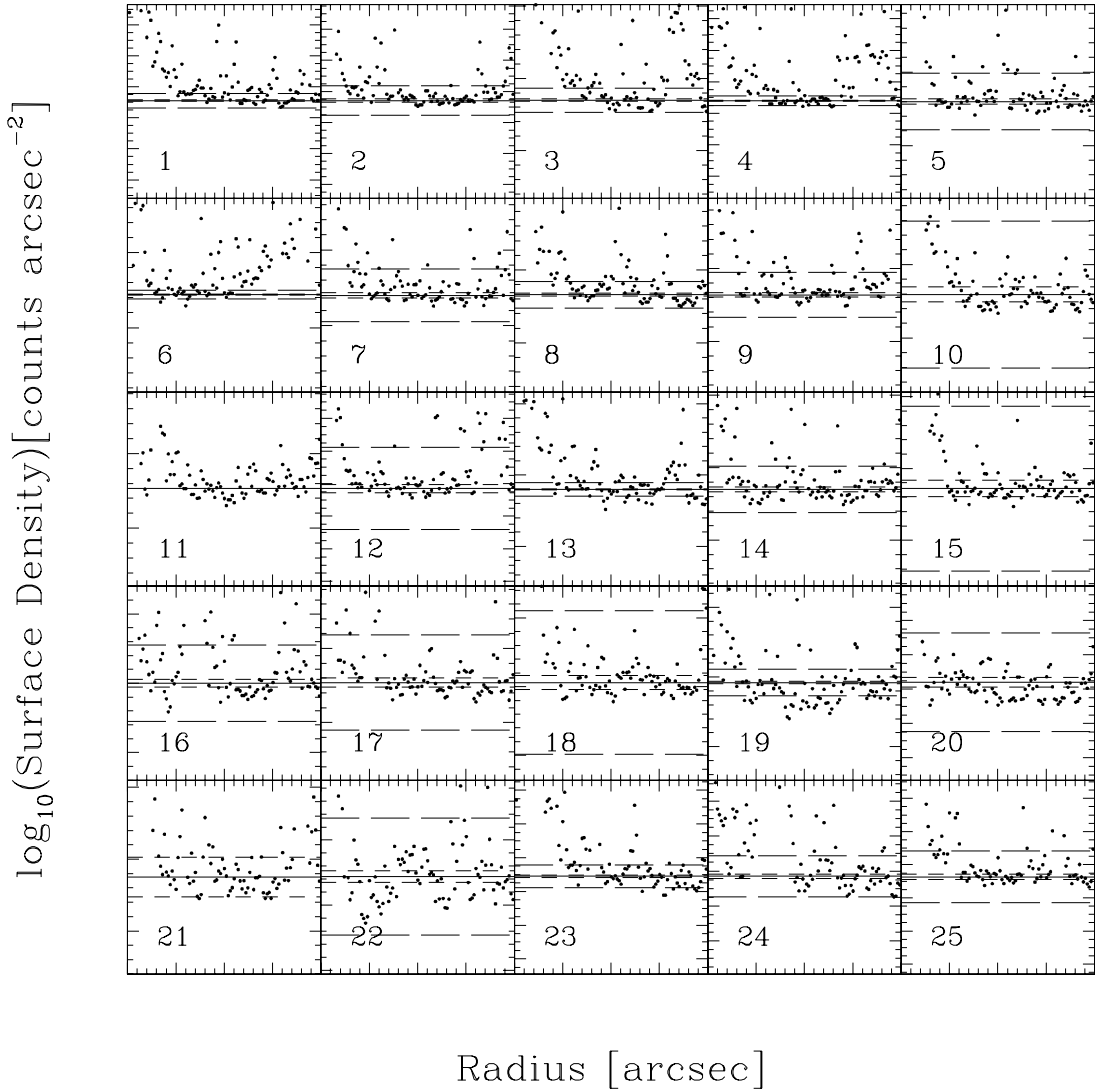}
\caption{Background fits for the first 25 LMC clusters. The range of the horizontal axis is 0$^{\prime \prime}$ to 200$^{\prime \prime}$ for all clusters. The range of the vertical axis is  $\pm$20$\%$ of the cluster's background value, and so is different for each cluster. The solid line marks the adopted background value, which is usually the median within the background annulus described in the text. The short and long dashed lines denote $\pm$1$\%$  and $\pm 10$\% of the central surface brightness. Sky uncertainties that are small relative to the central brightness of the cluster have little impact on the derived cluster parameters. All background fits are available on-line.}
\label{fig:Background Fits}
\end{center}
\end{figure*}

\begin{figure*}[htbp]
\begin{center}
\plotone{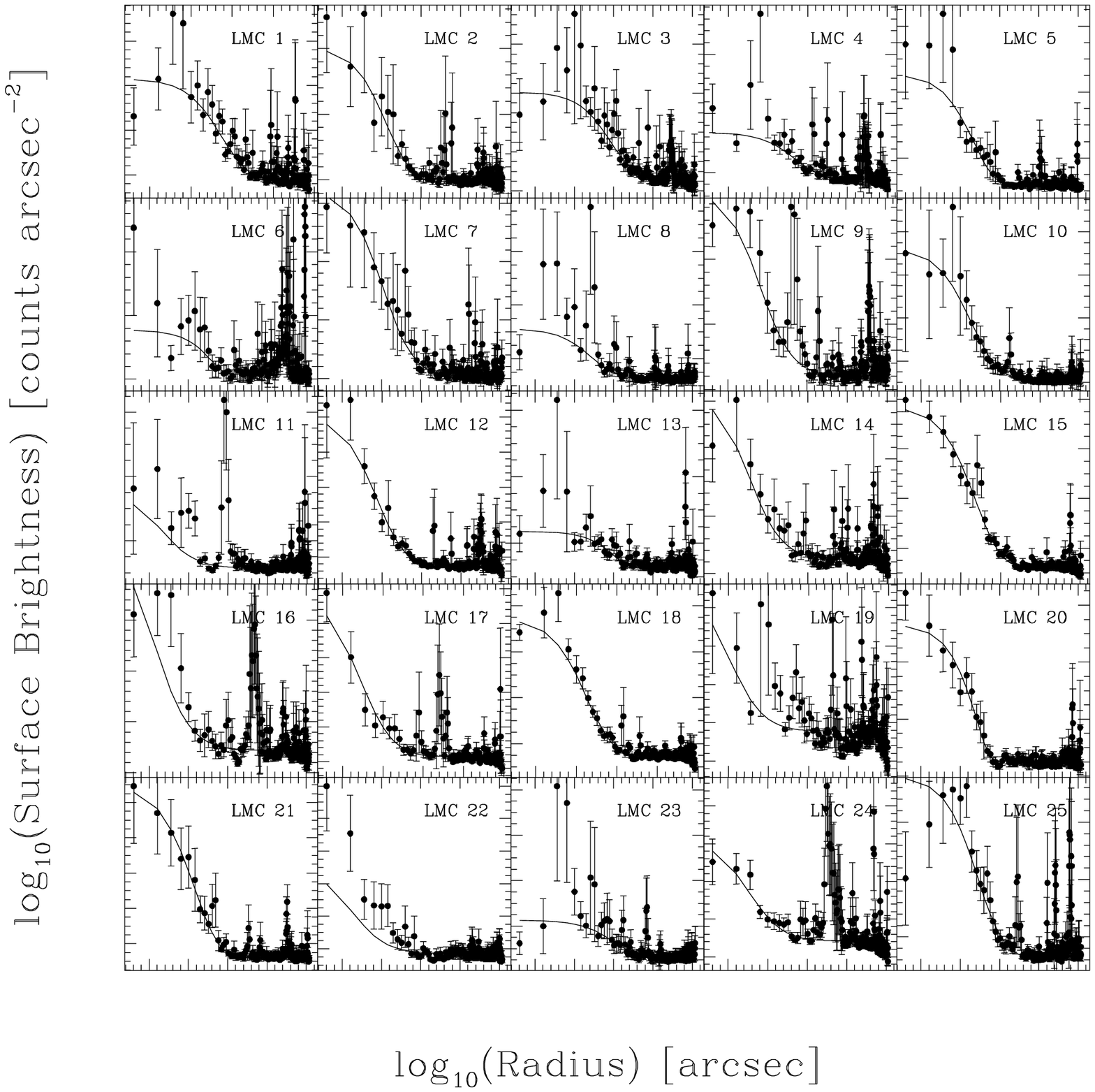}
\caption{$V$-band surface brightness profiles of the first 25 LMC clusters and King model plus background fits. The range of the horizontal axis is 0$^{\prime \prime}$ to 200$^{\prime \prime}$ for all clusters, while the vertical axis is set for each cluster. All cluster profiles are available on-line.}
\label{fig:Sample Profiles}
\end{center}
\end{figure*}

Our sample, like all samples, must be used with caution
in evolutionary studies because clusters of particular ages, masses, and/or surface brightnesses may be preferentially missing. Differences in assumptions regarding this 
incompleteness
have led to apparently conflicting results in the literature \citep{chandar,gieles,degrijs}, but correcting for the incompleteness is, as always, model dependent. Because the source photometric catalog is published, it is possible for subsequent investigators to recover the selection function by placing artificial clusters into the catalog and repeating the cluster selection. We do not attempt to correct our statistics here because the selection function
depends on assumptions regarding the parent population, and as such this work should be done by the individuals testing a particular model.

\section{Profile Fitting}

\begin{figure*}[htbp]
\begin{center}
\plotone{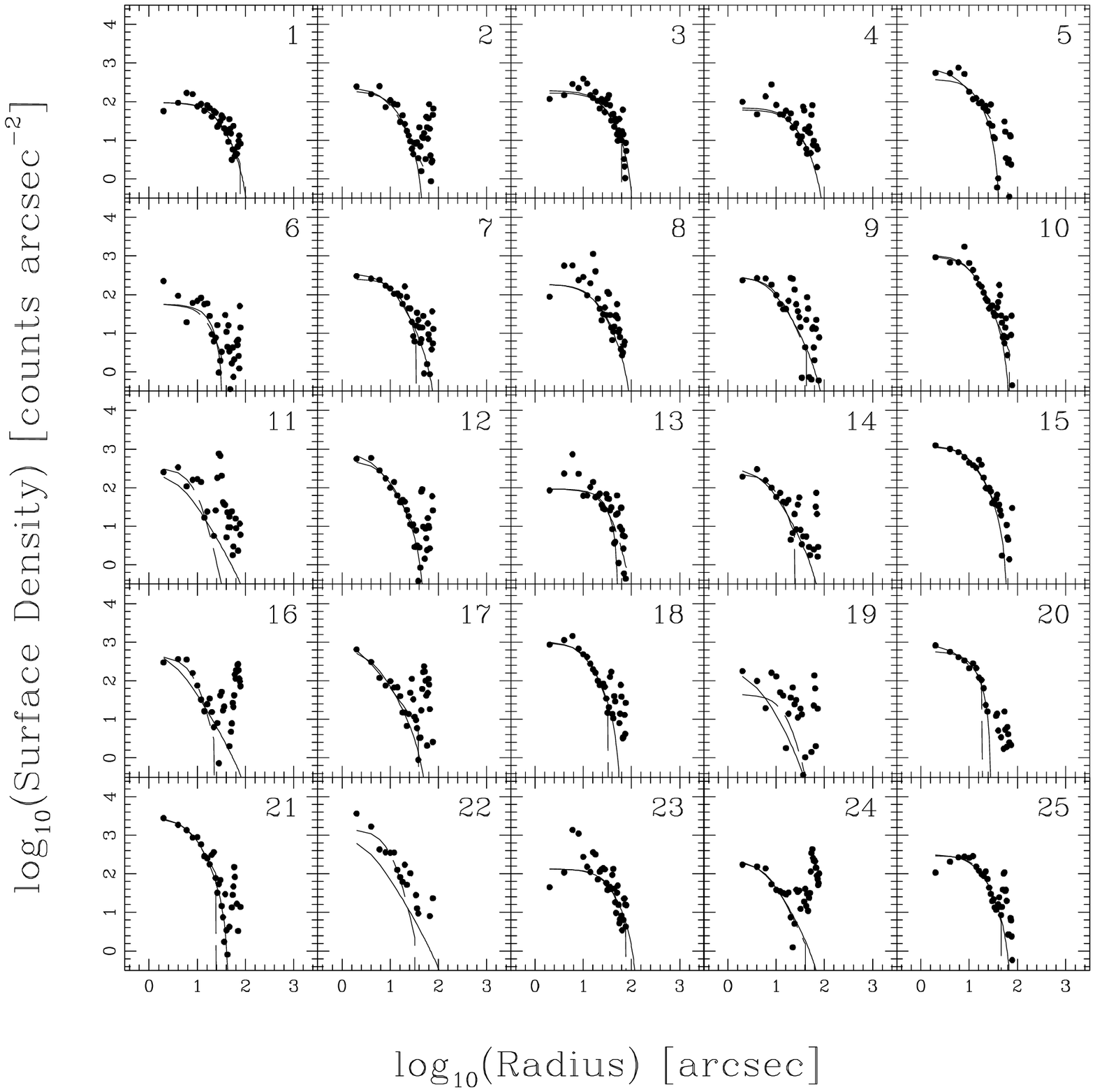}
\caption{Sky-subtracted surface brightness profiles in logarithmic units. Once the sky-subtracted profile drops below zero we terminate the plotting for that cluster. We overlay the best fit King (solid line) and EFF (long dashed line) profiles. All profiles are available on-line.}
\label{fig:Subtracted Sky Fits}
\end{center}
\end{figure*}

The process of profile fitting begins with the definition of the cluster center. We identify the center using the stellar overdensity as calculated from the photometric catalog. However,
because the catalog is grossly incomplete toward the
overcrowded cluster centers, the centers calculated from the catalog
often do not correspond precisely to the correct location, where the correct location is defined from visual inspection.
Instead, we return to the images themselves and recenter by calculating the luminosity
weighted centroid within 20$^{\prime\prime}$ of the initial estimated centroid. We reject the rare, brightest pixels (those that exceed the image mode by 3000 or more counts) from this calculation. 
We iterate the procedure seven times, each time using the new center to define the aperture. Because
even this process can sometimes be corrupted, we visually examine all candidate clusters
and interactively adjust those centers that are evidently incorrect. The cluster images presented in 
Figure \ref{fig:Sample Images} are centered on our final adopted centers, so the reader can judge the
fidelity of our centering process. For the poorer, irregular clusters the center is often ill-defined.

The second step in determining the cluster surface brightness 
profiles is the accurate determination of the
background flux. This step is particularly problematic in areas of high stellar density or nebulosity.
We initially calculate the background as the median within a 60 pixel wide (42 arcsec) annulus starting 180 pixels (126 arcsec) from the 
cluster center, which is larger than the tidal radius of all of our clusters. Because we do not know the cluster size prior to this procedure, and therefore whether we have defined an optimal background aperture, we iterate on the background determination, defining the beginning of the background annulus using the previous estimate of the tidal radius. The accurate determination of the background is a key step because adopting a background level that is $\pm10$\% off from the correct background often results in a poor fit \citep{hz}.

We fit the combination of model profile and background to the $V$-band luminosity profile as measured using circular apertures directly on the cluster images. Of the four bands available, $V$ is the deepest and typically has the best image quality. Although individual cluster stars are often highly confused in the clusters centers, the centers are far from saturated and so the luminosity density is a fair tracer of the stellar density (barring non-uniformity in the stellar radial mass function). We search for the best fit by examining models over a range of concentrations, $0.001 \le c \le 2$, central surface brightness that range from 0.25 to 1.75 the mean
central (4 arcsec) surface brightness, and core radii that range from 1 to 101 arcsec. Uncertainties in each parameter, presented as the high and low estimates in Tables \ref{tab:Structural Parameters of King model} and \ref{tab:Structural Parameters of EFF model}, are estimated by tracking all models that cannot be rejected with greater than 67\% confidence. If even the best fit
model is one that can be rejected with such confidence, then parameter uncertainties are not given.

\begin{figure*}[htbp]
\begin{center}
\plotone{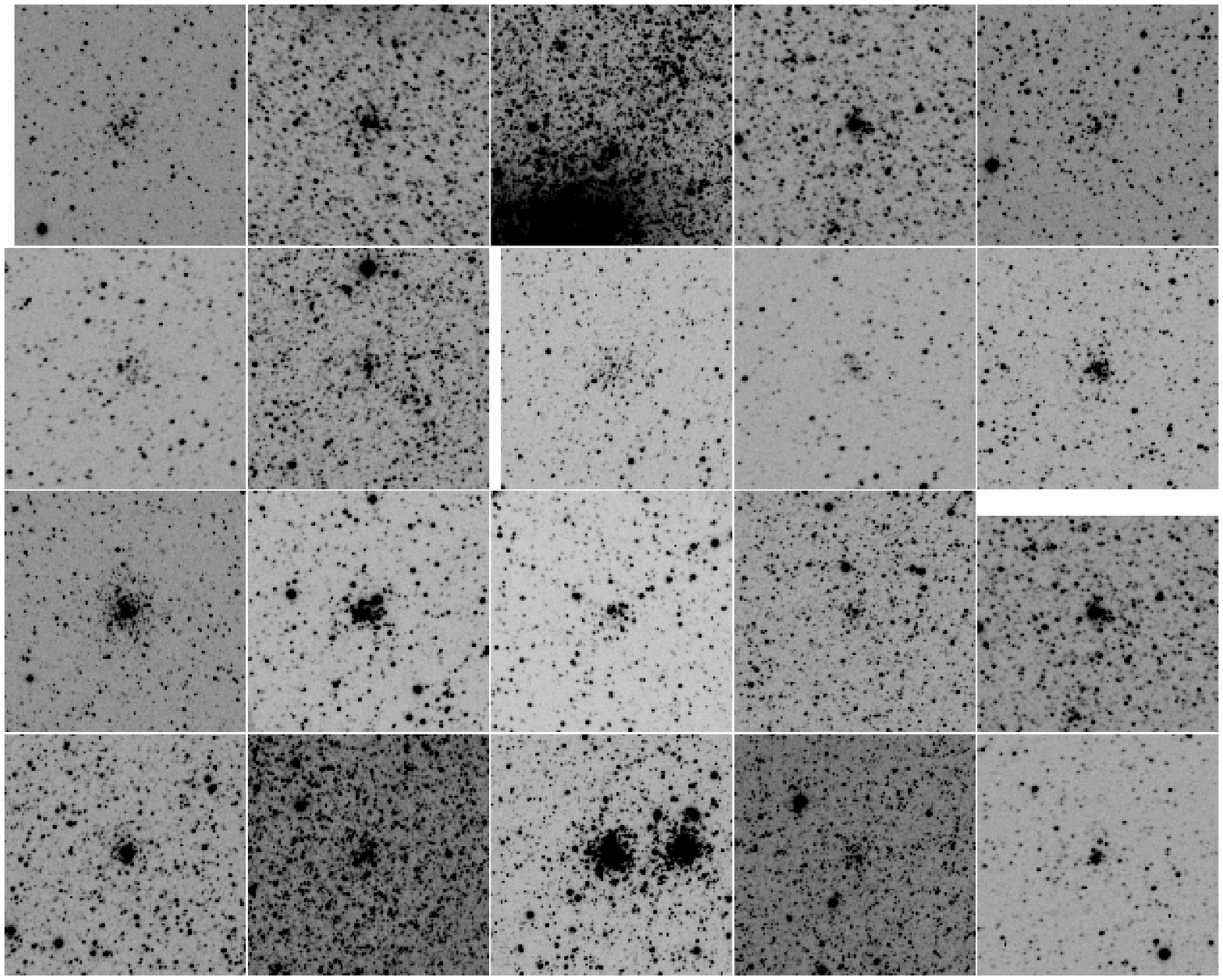}
\caption{$V$-band images of the 20 clusters with the largest $\chi^2_{\nu}$ values. From the upper left (largest $\chi^2_{\nu}$) these are clusters 191, 91, 539, 125, 365, 111, 150, 605, 132, 109, 69, 994, 844, 75, 127, 647, 347, 847, 520, and 116.}
\label{fig:High ChiSq}
\end{center}
\end{figure*}

For consistency with our previous work, the functional forms used to model the profiles are the \cite{king62} and EFF \citep{elson} profiles, as used by \cite{hz}.
\cite{mclaughlin} made the case in favor of a third functional form, the \cite{wilson} profile. In general,
we find that both the King and EFF profiles provide acceptable fits to our candidate clusters because the differences between the model profiles lie primarily at large radii, where our S/N is poor and background uncertainties dominate. We expect the same to be true if we were to compare to the Wilson profile. Detailed 
comparisons between profiles for particular clusters should be done with deeper, higher resolution data, such as that from the {\sl Hubble Space Telescope} \citep{mclaughlin} and we confine ourselves to a rough,
statistical comparison of the King and EFF profiles.

The tidal radius, the radius at which stars become unbound to the cluster and in practice
where the projected density drops to zero, is the structural parameter that is most sensitive to the adopted background level. We find that our uncertainty in the background often results in a
tidal radius whose uncertainty is so large as to make the measurement practically meaningless. Instead, we choose to 
present a more robust measure of the size of the cluster, the radius that includes 90\% of the light, $r_{90}$, calculated
from the fitted model profile.  In addition to the iterative calculation, we visually examine all the background determinations to determine if the fit may have been contaminated and manually adjust as necessary. In Figure \ref{fig:Background Fits} we present the background fits. The Figure and Table \ref{tab:Structural Parameters of King model} present the final adopted background values. The most likely source of error in the determination of the profiles for most clusters is
in the determination of the background level. The Figure enables the reader to examine the definition of the background for each cluster.

\begin{figure}[htbp]
\begin{center}
\plotone{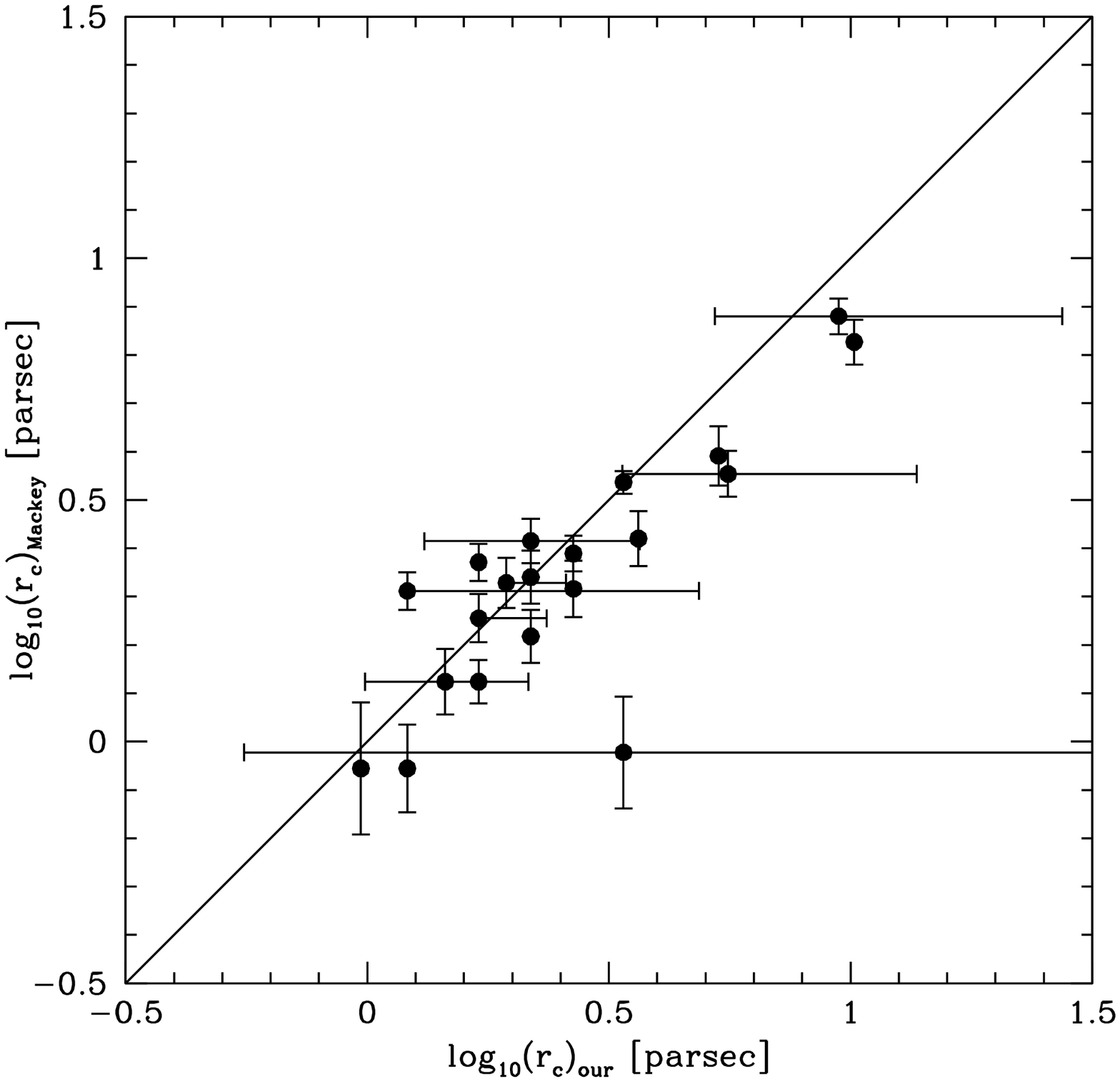}
\caption{We comapre our EFF core radius values for cluster matches from \cite{mackey03b}. The clusters for which we have matches are 142, 277, 816, 207, 734, 970, 917, 513, 483, 593, 1055, 346, 122, 67, 903, 999, 415, 546, 35, and 1063}
\label{fig:Mackey Core Radius}
\end{center}
\end{figure}

\begin{figure}[htbp]
\begin{center}
\plotone{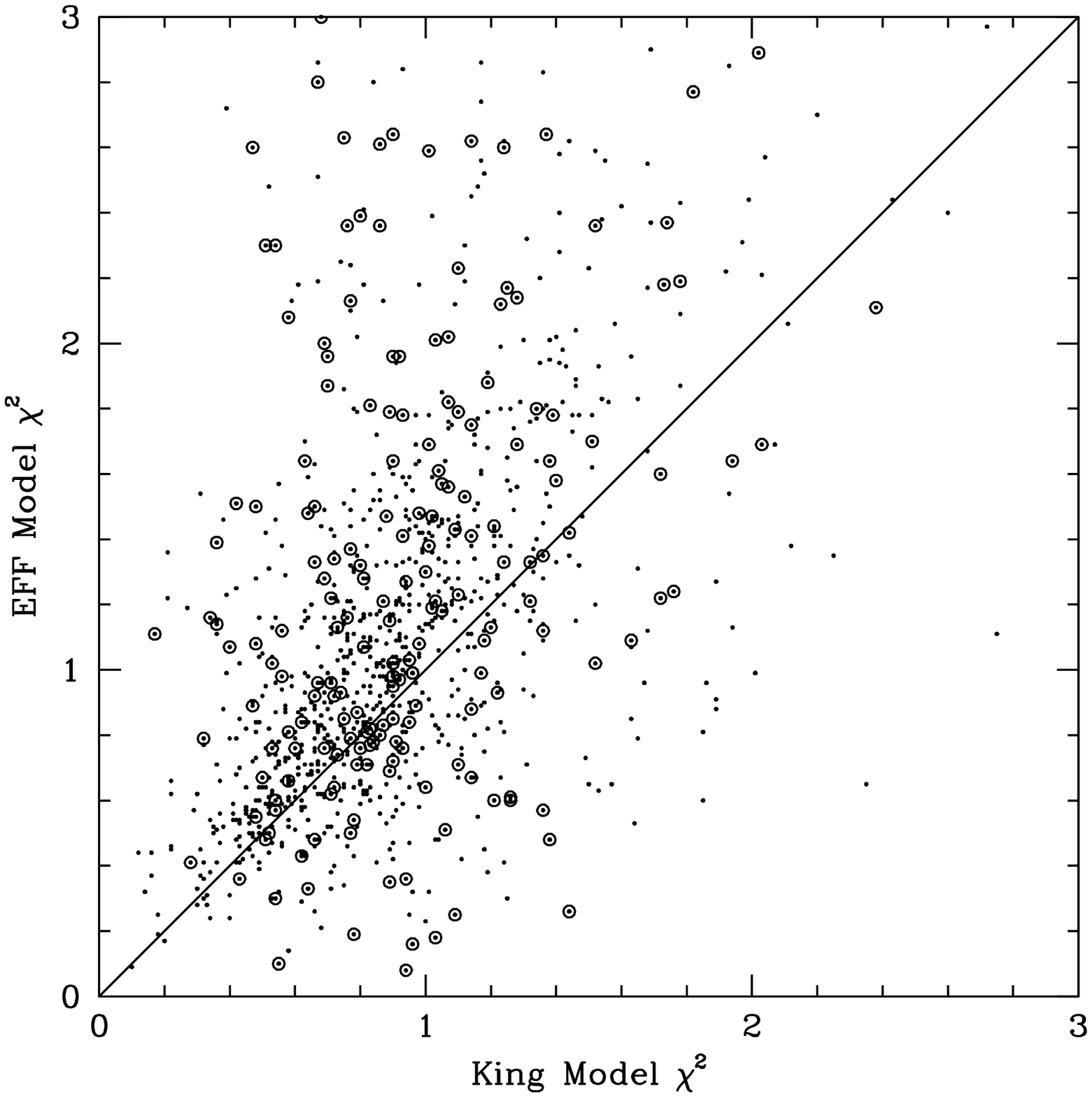}
\caption{Comparison of $\chi^2_{\nu}$ values for King and EFF model fits. We have excluded clusters with statistically unacceptable fits (those that can be rejected with $>$ 90\% confidence). The solid line is the 1:1 line. The clusters where the central surface brightness is greater than 10 times the sky value are shown with an open circle around the point.
EFF $\chi^2_{\nu}$ values are systematically larger than those from King model fits, although not 
exclusively, suggesting that perhaps for some sub-class of cluster the EFF models are superior.}
\label{fig:Chi Squared Plot}
\end{center}
\end{figure}

Of the 1066 clusters in our final catalog, 269 ($\sim$ 25\%) are discrepant with the King profile at 67\% confidence, which is in accordance with the expectation due to random fluctuations and we  conclude that the bulk of the sample is well described by King models given the available data. 
The reduced chi-squared values, $\chi^2_{\nu}$, for the King models are somewhat 
lower than those for the EFF, suggesting, as was the case for SMC clusters \citep{hz}, that the King
models provide a better basis set for the general cluster sample. Of course, specific subsamples, for example younger clusters, may be better fit by the EFF models \citep[as originally proposed by][]{elson}. 

\begin{figure}[htbp]
\begin{center}
\plotone{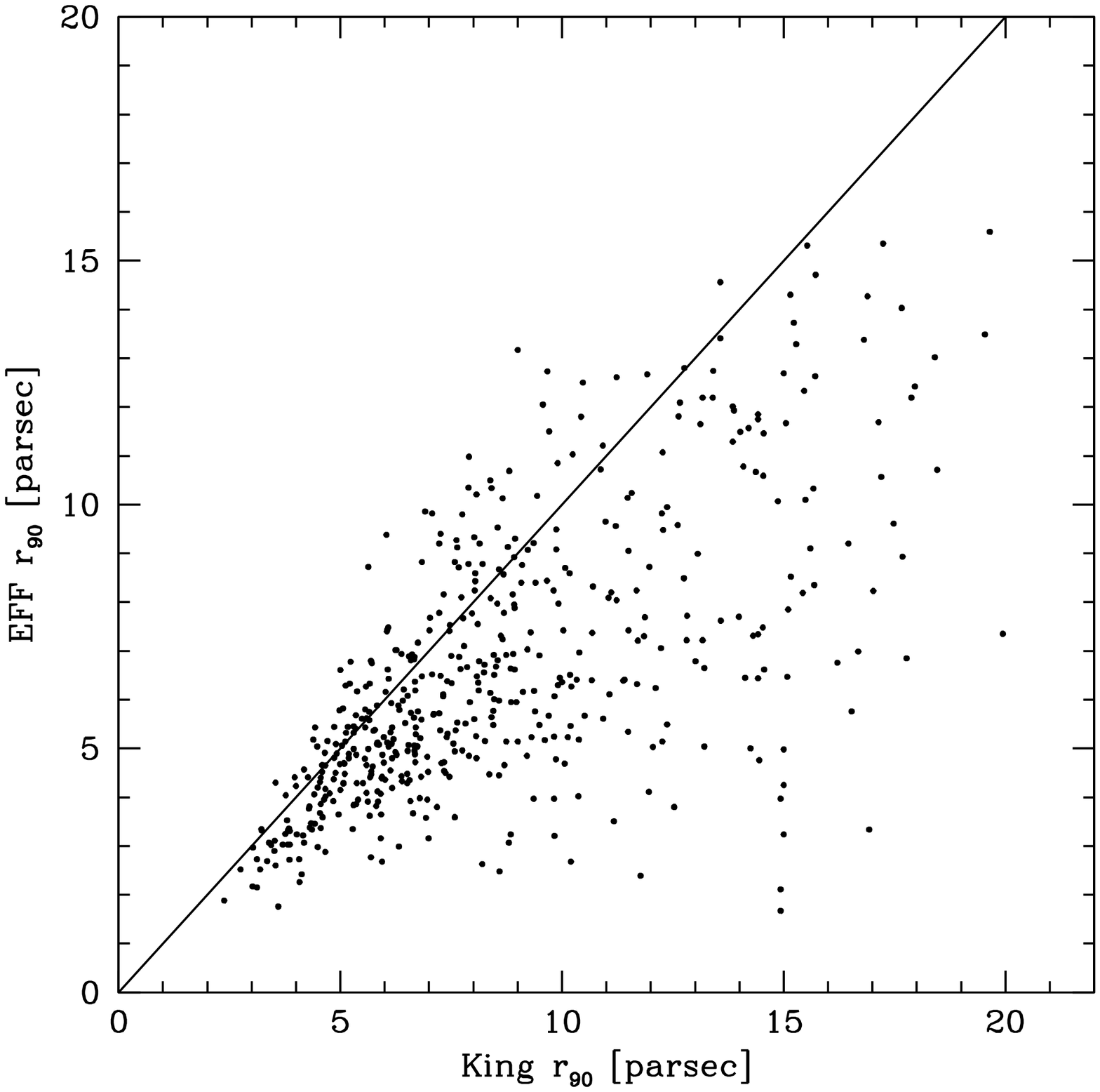}
\caption{Comparison of $r_{90}$ values for King and EFF model fits. The solid line is the 1:1 line.}
\label{fig:r90 Plot}
\end{center}
\end{figure}

\begin{figure}[htbp]
\begin{center}
\plotone{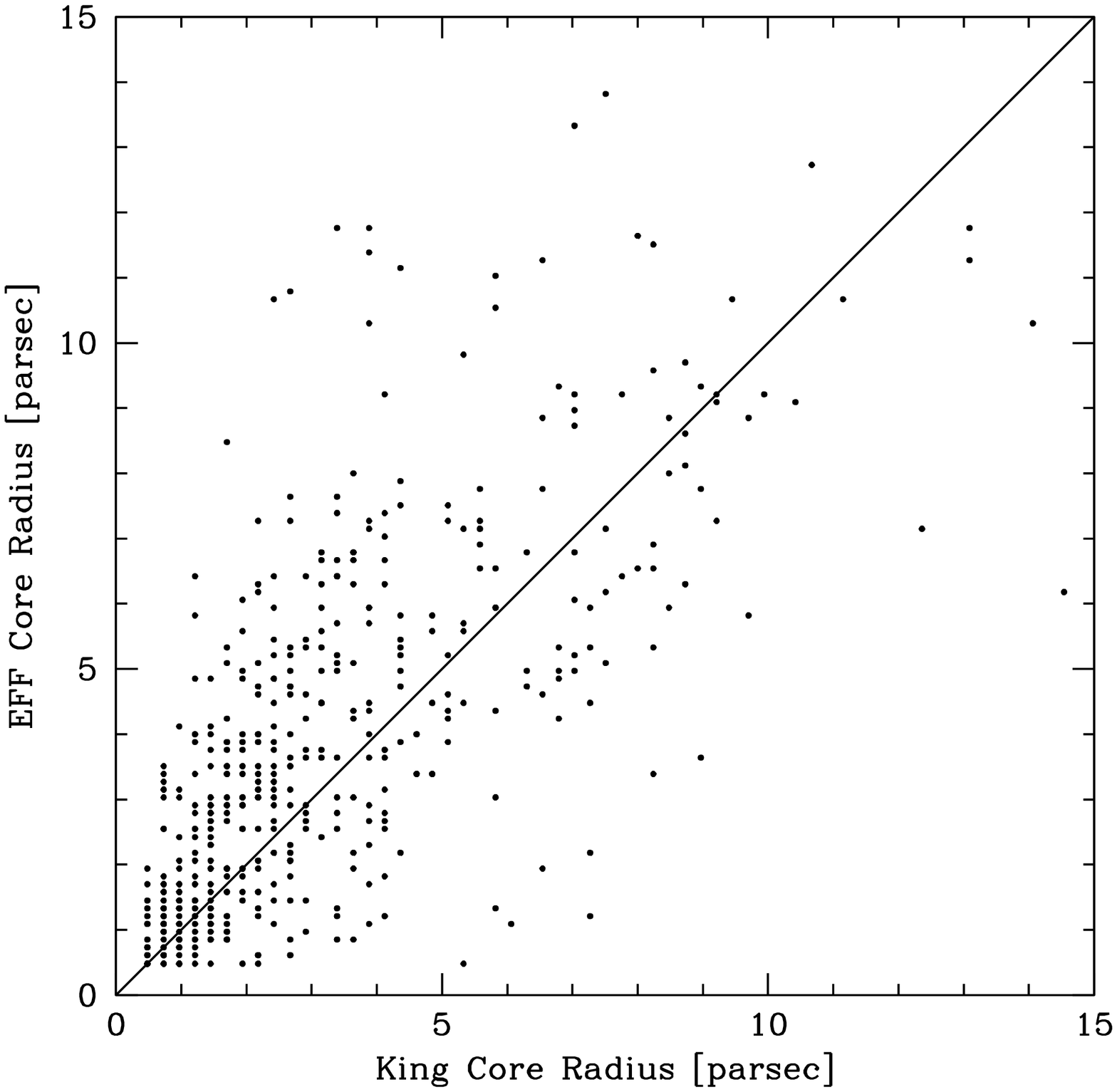}
\caption{Comparision of $r_{c}$ values for King and EFF model fits. The solid line is the 1:1 line.}
\label{fig:cr Plot}
\end{center}
\end{figure}

In summary, we provide the cluster coordinates and cross-identifications in Table \ref{tab:Names and Positions},  the best fit parameters in Table \ref{tab:Structural Parameters of King model}, 
the sky region and fits in Figure \ref{fig:Background Fits}, the luminosity profiles plus background and fits in Figure \ref{fig:Sample Profiles},  and the sky-subtracted profiles in Figure \ref{fig:Subtracted Sky Fits}. In Table \ref{tab:Structural Parameters of King model} we present, in matching order to the listing in 
Table \ref{tab:Names and Positions}, the low, best, and high estimates of the core radii in pc (1 $\sigma$ uncertainties), the low, best, and high estimates of the 90\% enclosed radii in pc, the integrated V magnitude based on the model fits, the central surface brightness in counts/sq. arcsec as
measured from the models, the background value in counts/sq. arcsec, and the value of $\chi^2_{\nu}$.
We examine the profile plots for all of the clusters to determine if any needed to be refit with slightly
different center or sky values. We refit roughly 40\% with slight changes in centroid or background
value, but usually the differences in the fit parameters are negligible. 

\begin{deluxetable*}{crrrrrrrrrr}
\tablecaption{Structural Parameters for King Model Fits}
\tablewidth{0pt}

\tablehead{
\colhead{Cluster} &
\multicolumn{3}{c}{Core Radius (pc)} & \multicolumn{3}{c}{90\% Light Radius (pc)} & \colhead{m$_V$} & \colhead{$\Sigma_0$} & \colhead{$\chi^{2}_{\nu}$} & \colhead{BKG}\\
\colhead{    } & 
\colhead{low}   & 
\colhead{best}   &
\colhead{high} &
\colhead{low} &
\colhead{best} & 
\colhead{high} &
\colhead{} &
\colhead{counts arcsec$^{-2}$} &
\colhead{} &
\colhead{}  
}

\startdata

1  & $3.64$  & $5.58$  & $11.15$         & $11.79$  & $15.53$  & $17.52$     & $13.9$    & $94$   & $0.83$ & $311$\\
2  & $0.48$  & $2.67$  & $9.70$          & $5.28$   & $6.84$   & $21.63$     & $15.0$    & $190$  & $0.60$ & $314$\\
3  & $3.15$  & $7.03$  & $21.33$         & $10.87$  & $15.00$  & $19.10$     & $13.1$    & $168$  & $0.66$ & $334$\\
4  & $2.67$  & $5.33$  & $9.45$          & $10.54$  & $13.89$  & $18.46$     & $14.6$    & $61$   & $0.81$ & $308$\\
5  & $1.21$  & $3.39$  & $10.42$         & $4.66$   & $6.15$   & $8.46$      & $14.1$    & $385$  & $0.69$ & $330$\\
6  & $0.48$  & $7.76$  & $11.15$         & $5.92$   & $8.74$   & $23.57$     & $15.7$    & $56$   & $0.53$ & $322$\\
7  & $0.97$  & $2.18$  & $10.42$         & $5.46$   & $10.33$  & $17.19$     & $14.2$    & $347$  & $0.54$ & $320$\\
8  & $0.48$  & $3.64$  & $11.15$         & $6.56$   & $12.63$  & $20.14$     & $14.0$    & $88$   & $0.52$ & $333$\\
9  & $0.73$  & $1.70$  & $4.85$          & $4.47$   & $11.87$  & $19.37$     & $14.6$    & $63$   & $0.57$ & $331$\\
10 & $0.48$  & $2.42$  & $11.88$         & $5.37$   & $8.41$   & $20.05$     & $13.1$    & $1217$ & $0.42$ & $329$\\

\enddata

\tablecomments{The complete version of this Table is available in the electronic edition of
the Journal. The printed edition contains only a sample.}
\label{tab:Structural Parameters of King model}
\end{deluxetable*}

\begin{deluxetable*}{ccccccccc}
\tablecaption{Structural Parameters for EFF Model Fits}
\tablewidth{0pt}

\tablehead{
\colhead{Cluster} &
\multicolumn{3}{c}{Core Radius (pc)} & \multicolumn{3}{c}{90\% Light Radius (pc)} & \colhead{Central Surface Brightness} & \colhead{$\chi^{2}_{\nu}$}\\
\colhead{    } & 
\colhead{low}   & 
\colhead{best}   &
\colhead{high} &
\colhead{low} &
\colhead{best} & 
\colhead{high} &
\colhead{counts arcsec$^{-2}$} &
\colhead{}  
}

\startdata

1  &  $3.03$ &   $6.91$ &  $14.67$ &  $12.53$ &  $15.31$ &  $16.71$   &    $94$ &  $0.85$\\
2  &  $0.48$ &   $2.30$ &   $8.73$ &   $5.11$ &   $8.82$ &  $10.49$   &   $226$ &  $0.51$\\
3  &  $2.55$ &   $6.79$ &  $16.12$ &  $11.38$ &  $12.69$ &  $13.42$   &   $190$ &  $0.84$\\
4  &  $0.00$ &   $6.06$ &   $0.00$ &   $0.00$ &   $9.97$ &   $0.00$   &    $68$ &  $1.17$\\
5  &  $0.73$ &   $1.21$ &   $8.00$ &   $5.04$ &   $5.93$ &   $6.18$   &   $756$ &  $0.61$\\
6  &  $0.48$ &   $6.42$ &   $8.61$ &   $4.96$ &   $6.92$ &   $8.23$   &    $56$ &  $0.78$\\
7  &  $1.58$ &   $6.30$ &   $7.64$ &   $5.67$ &   $6.41$ &   $7.07$   &   $256$ &  $0.77$\\
8  &  $0.97$ &   $4.36$ &  $12.12$ &   $6.72$ &  $11.81$ &  $13.14$   &   $183$ &  $0.67$\\
9  &  $1.33$ &   $2.67$ &   $5.58$ &   $5.11$ &   $7.69$ &   $8.82$   &   $287$ &  $0.89$\\
10 &  $0.61$ &   $2.91$ &  $11.76$ &   $4.92$ &  $10.34$ &  $14.51$   &  $1071$ &  $0.41$\\

\enddata

\tablecomments{The complete version of this Table is available in the electronic edition of
the Journal. The printed edition contains only a sample.}
\label{tab:Structural Parameters of EFF model}
\end{deluxetable*}

\begin{figure*}[htbp]
\begin{center}
\plotone{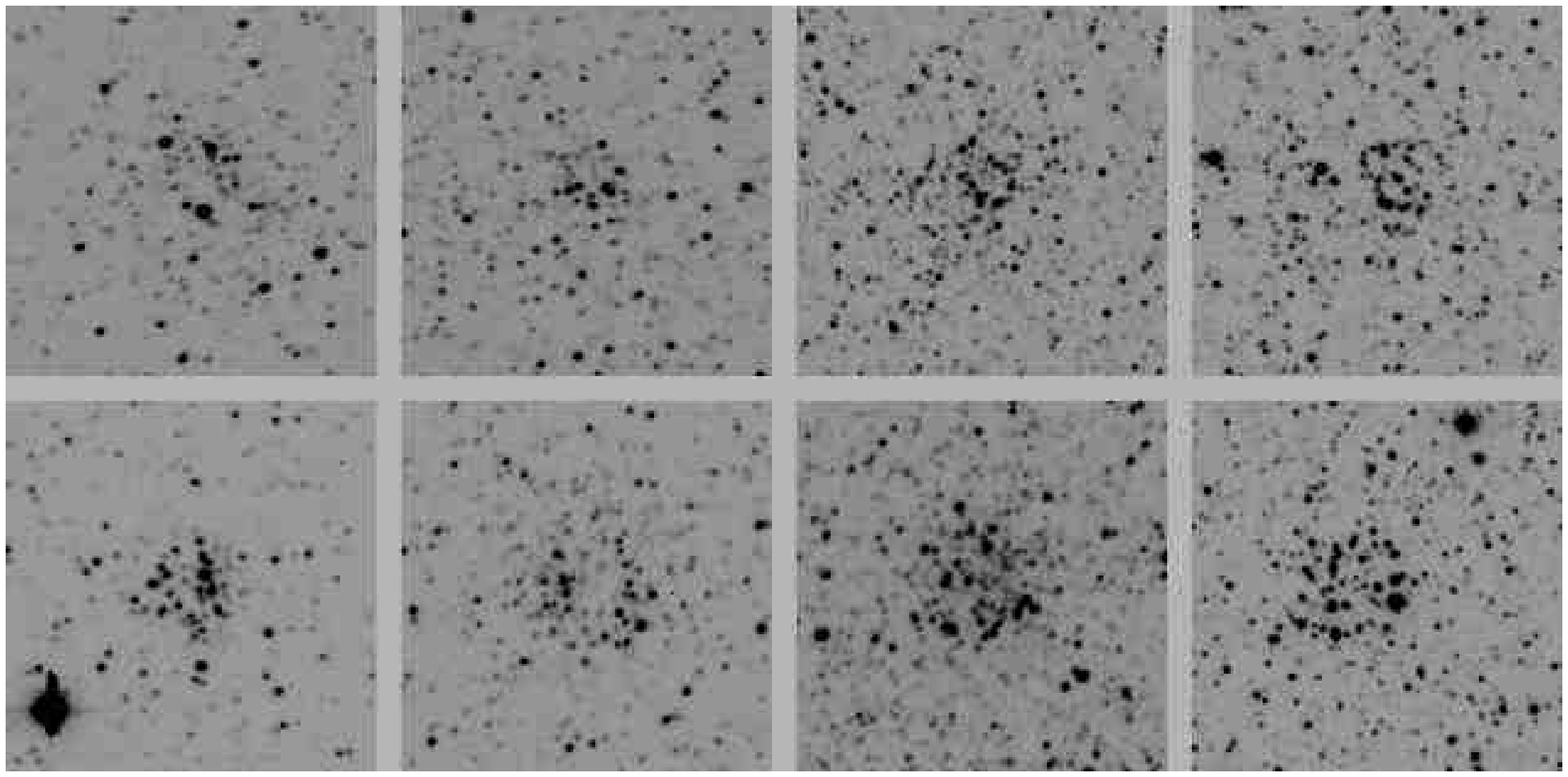}
\caption{$V$-band images of 8 example ring clusters. From the upper left these are 23, 77, 199, 233, 241, 632, 693, and 1042. Images sizes correspond to 105 arcsec on a side.}
\label{fig:Ring Clusters}
\end{center}
\end{figure*}

In Figure \ref{fig:High ChiSq} we show the $V$-band
images of the 20 clusters with the highest $\chi^2_{\nu}$ values. This Figure demonstrates
that poorly fitting clusters are not necessarily extremely poor or marginal clusters. On occasion they have a neighboring cluster that affects the fit, but in most cases the appear normal. This suggests that the high $\chi^2_{\nu}$ reflects a more basic problem with the fitting profile rather than with the data themselves. However, it is also true that in such a large sample, one naturally expects some outliers, particularly because our statistical errors do not include for the possibility 
background fluctuations.

We test our EFF-model fit parameters against those derived using superior data presented by 
\cite{mackey03b}. They do not present $r_{90}$ for their clusters and we do not trust our estimates of the tidal radii, so the one radius we can compare between the two studies is the core radius, $r_c$. There are 16 clusters in common with the necessary data between the two studies and the comparison is presented in Figure \ref{fig:Mackey Core Radius}.
The agreement is generally good, although there may be an indication for a slight systematic
bias in the sense that we would be overestimating $r_c$. Such a bias would not be surprising given the superior resolution of 
the {\sl HST} data, although if we consider only those clusters for which our fits 
cannot be rejected with $>$ 90\% confidence (those with x axis error bars) we find weaker evidence for any systematic difference between the two studies.

\section{Profile Characteristics}

\subsection{Comparison of Different Model Profiles}

Both King and EFF profiles fit most of the clusters well. 
However, as \cite{hz} found for the SMC clusters, the EFF fit is 
generally associated with a higher $\chi^2_{\nu}$ value (Figure \ref{fig:Chi Squared Plot}).
We exclude from this plot the clusters where either King or EFF profiles result in a statistically unacceptable fit ($> 90$\% confidence of rejection). The most luminous clusters are denoted with open circles (central surface brightness $>$ 10 times its sky value) and these illustrate that the difference in fit statistics is not limited to low signal-to-noise clusters. Nevertheless,
the differences can be rather subtle (see Figure \ref{fig:Subtracted Sky Fits}) and one question
is whether the choice of model greatly affects the derived parameters one might want to use
to quantify cluster structure, such as the core radius or size. 
In Figure \ref{fig:r90 Plot}, we compare the $r_{90}$ values for the King and EFF models and find that the EFF 
$r_{90}$ tends to have a larger $r_{90}$ value than King. For the smaller clusters the differences
appear modest ($\sim 25$\%) but the values are correlated. However, for the larger clusters ($r_{90} > 30$ arcsec) the correlation breaks down and the difference can amount to factors of several. This 
differences arises from the extrapolation of the profile to radii larger than those for which our
data provide good constraints and the implication of significant amounts of light at large radius, which
of course drives $r_{90}$ upwards. Due to the truncated nature of King profiles, we prefer those
in determining $r_{90}$ because there is less potential for those profiles to ``run away" at large
radii. The differences between the models are less marked when comparing estimates of $r_c$ (Figure \ref{fig:cr Plot}).

\subsection{Ring Clusters}

\begin{figure}[htbp]
\begin{center}
\plotone{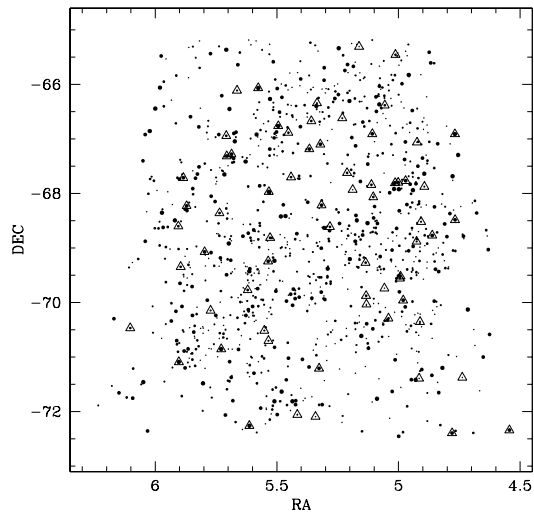}
\caption{The  distribution of all the LMC clusters according to their RA and DEC measurements. Dot size scales with core radius and ring clusters are additionally designated
by triangles. There is no evident connect in cluster properties with their spatial distribution.}
\label{fig:LMC dist}
\end{center}
\end{figure}

In the SMC, \cite{hz} had noted a small, but interesting sub-class of clusters that appear ring-like.
We find similar clusters in the LMC and show such clusters in 
Figure \ref{fig:Ring Clusters}. From our visual inspection of the profiles (Figure \ref{fig:Sample Profiles}, we classify 78 as ring clusters ($\sim 7$\% of our sample). The nature of these clusters remains
a mystery that detailed kinematics might help resolve. We plot their projected distribution, as well
as all other clusters, in the LMC in Figure \ref{fig:LMC dist} but find no telltale difference 
in the distribution of the ring clusters. 

\subsection{Comparison to the Cluster Population of the Small Magellanic Cloud}

\begin{figure}[htbp]
\begin{center}
\plotone{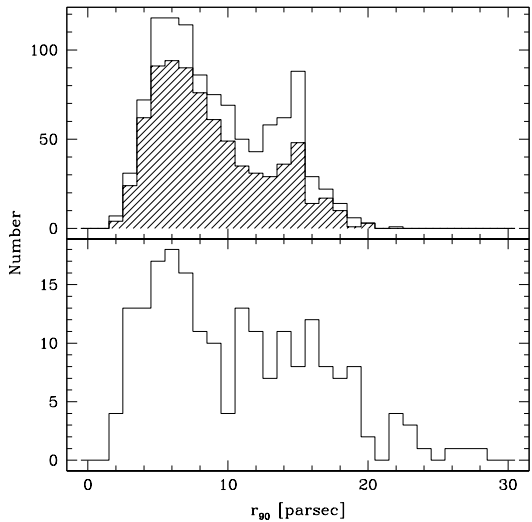}
\caption{Comparison of the $r_{90}$ values for clusters in the LMC (top panel) and SMC (bottom panel). Values come from our King model fits. Despite having a factor of five fewer clusters, the SMC has two times as many clusters with $r_{90} > 20$ pc. The shaded portion of the upper histogram represents the clusters with acceptable fits (those that cannot be rejected with $>$ 90\% confidence)} determined by the $\chi^2_{\nu}$ value.
\label{fig:R90 Comparison}
\end{center}
\end{figure}

We compare the profile structural parameters, more specifically $r_{90}$, core radius and cluster concentration, for our LMC and the published SMC cluster populations \citep{hz}. In Figure \ref{fig:R90 Comparison} we compare the $r_{90}$ values. We find that the distributions are very similar for $r_{90} < 15$ pc, but that there is a significant drop in the fraction of  clusters with $r_{90} > 15$ pc in the LMC relative to the SMC. The tail of large clusters extends in the SMC, even though the SMC has one-fifth the total number of clusters. We confirm
this visual impression using a KS test to calculate that the likelihood of these two populations being drawn from the same parent distribution is $9 \times 10^{-5}$.  Similarly, \cite{bica} find that the size distribution of associations, in particular, are more populated at large sizes in the SMC than in the LMC (see their Figure 4).

\begin{figure}[htbp]
\begin{center}
\plotone{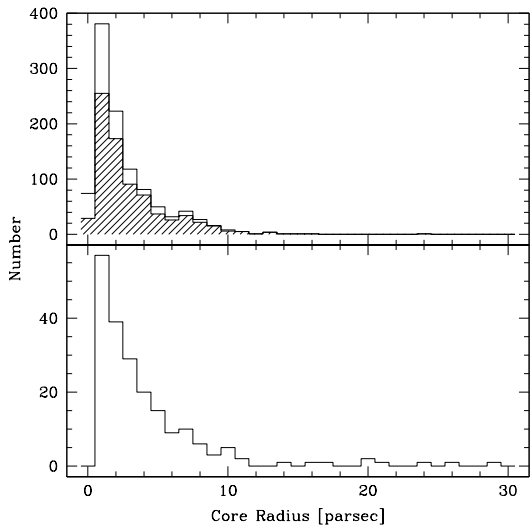}
\caption{Comparison of the $r_c$ values in the LMC (top panel) and SMC (bottom panel). Values come from the King model fits. The shaded portion of the upper histogram represents the clusters with acceptable fits (those that cannot be rejected with $>$ 90\% confidence) determined by the $\chi^2_{\nu}$ value.}
\label{fig:Core Radius Comparison}
\end{center}
\end{figure}

\begin{figure}
\begin{center}
\plotone{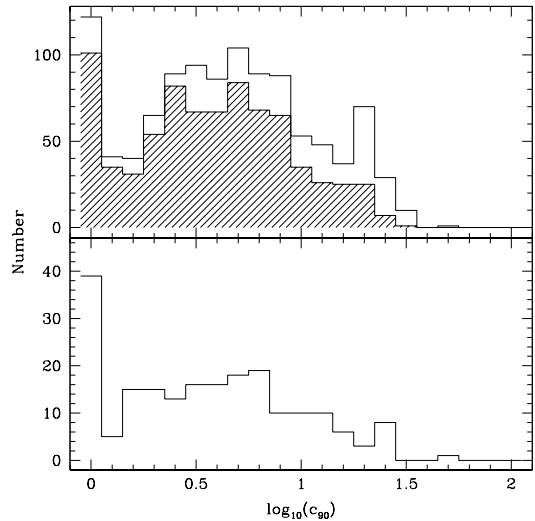}
\caption{Comparison of the concentration values in the LMC (top panel)  and SMC (bottom panel).
 The concentration is calculated
in a non-standard manner in that we use $r_{90}$ rather than the tidal radius and denoted using $c_{90}$. This is done
because our tidal radii are often quite uncertain. The values used correspond to those
from our King model fits. The shaded portion of the upper histogram represents the clusters with statistically acceptable fits (those that cannot be rejected with $>$ 90\% confidence) determined by the $\chi^2_{\nu}$ value.
}
\label{fig:Concentration Comparison}
\end{center}
\end{figure}

There is the potential that this result is related to an unknown selection bias rather than to
a physical effect. Because the systems are found in the SMC and not in the LMC, a likely suspect 
is the higher stellar density found in the LMC. Over the bulk of the survey area, the stellar densities
are not grossly different between the two galaxies, but the bar region of the LMC greatly exceeds
anything found in the SMC. If these larger systems tend to be found near the centers of galaxies
then we may have preferentially missed them in the LMC.  However, in the SMC they are found
throughout our survey region with a preference for the lower density, SMC wing. Our LMC survey
extends to low surface density regions, where we would expect to have detected such systems. Furthermore, the \cite{bica} compilation extends even further in radius around the LMC and should have also found these systems.

We follow-up on this finding by comparing the distributions of $r_c$. Again, we find that the bulk of the populations look quite similar, but that the SMC has a more significant tail of objects with large values, despite the smaller number of total clusters (Figure \ref{fig:Core Radius Comparison}).
Again, a KS test confirms the visual impression and the corresponding likelihood for the null
hypothesis is $2\times 10^{-8}$. Interestingly, the distribution of concentrations (Figure \ref{fig:Concentration Comparison}) are much more similar between the two cluster populations and they are not distinguishable at a $> 3\sigma$ level. We conclude that this population of clusters is larger in overall scale, not just outer envelope. We speculate that such clusters formed and/or survived preferentially in the SMC due to lower tidal stresses. Mass estimates would help confirm this hypothesis by enabling us to determine whether these are simply larger clusters in physical extent but not in mass, or if they are larger both in mass and size. 

\section{Summary}

We present a catalog of structural parameters for 1066 stellar clusters in the Large Magellanic Clouds. By construction, the catalog is of the same format as that published for the population of clusters in the Small Magellan Cloud \citep{hz}. Images, profiles, and the calculated parameters are presented for all clusters. As done for the SMC clusters, we intend to estimate ages
for these clusters in subsequent work to search for evolutionary signatures.

We find that \cite{king} and \cite{elson} profile fits, are statistically good fits for the bulk of the sample and that our parameters are in good agreement with existing data that is superior in quality but far more limited in quantity \citep{mackey03b}. One subset of clusters for which the fits fail are clusters that are underdense in their geometric centers, which we refer to as ``ring" clusters and which we also found in the SMC \citep{hz}. 

We find that the LMC cluster populations lacks a subpopulation of large clusters seen in the SMC. We find the lack of such clusters when comparing values of either $r_{90}$ (the radius that encloses 90\% of the luminosity) or $r_c$ (the King model core radius). In both cases, the likelihoods that the distributions are drawn from the same parent sample are $< 10^{-4}$. We speculate that the differences may be the result of a lower tidal field in the SMC, although cluster masses are needed to pursue this hypothesis further.

\end{document}